\documentclass[pre,twoside,twocolumn,superscriptaddress,floatfix]{revtex4-1}

\usepackage{graphicx,amssymb,amsmath,epsfig,color,enumerate,bm,psfrag}
\usepackage{etoolbox} 
\usepackage{afterpage,placeins,float}

\makeatletter
\appto{\appendix}{%
  \@ifstar{\def\theequation@prefix{A.}}%
          {}%
}
\makeatother
\epsfclipon

\begin{document}
 \title{Quenched trap model on the extreme landscape: \\the rise of subdiffusion and non-Gaussian diffusion}
 \author{Liang Luo}
 \affiliation{Department of Physics, Huazhong Agricultural University, Wuhan 430070, China}
 \affiliation{Institute of Applied Physics, Huazhong Agricultural University, Wuhan 430070, China}
 \author{Ming Yi}
 \email{yiming@mail.hzau.edu.cn}
 \affiliation{School of Mathematics and Physics, China University of Geosciences, Wuhan 430074, China}

\begin{abstract}
Non-Gaussian diffusion has been intensively studied in recent years, which reflects the dynamic heterogeneity in the disordered media. The recent study on the non-Gaussian diffusion in a static disordered landscape suggests novel phenomena due to the quenched disorder. In this paper, we further investigate the random walk on this landscape under various effective temperatures $\mu$, which continuously modulate the dynamic heterogeneity. We show in the long time limit, the trap dynamics on the landscape is equivalent to the quenched trap model, in which subdiffusion appears for $\mu<1$. The non-Gaussian distribution of displacement has been analytically estimated for short $t$, of which the stretched exponential tail is expected for $\mu\neq1$. Due to the localization in the ensemble of trajectory segments, an additional peak arises in $P(x,t)$ around $x=0$ even for $\mu>1$. Evolving in different time scales, the peak and the tail of $P(x,t)$ are well split for a wide range of $t$. This theoretical paper reveals the connections among the subdiffusion, non-Gaussian diffusion, and the dynamic heterogeneity in the static disordered medium. It also offers an insight on how the cell would benefit from the quasi-static disordered structures. 
\end{abstract}


\maketitle

\section{Introduction}

Dynamic heterogeneity \cite{berthier11,berthier11rmp,kirkpatrick15} has been recognized as the key feature of glassy systems, 
which refers to the widely spanned relaxation time of the disordered structures, the highly intermittent particle dynamics, and the large trajectory-to-trajectory fluctuations. The non-Gaussian diffusion, of which the distribution of particle displacement is not Gaussian, is observed in a wide range of disordered systems with dynamic heterogeneity,  including the crowding intracellular environments \cite{parry14,munder16,he16,jeon16}, colloidal\cite{wang09,sentjabrskaja16} and granular \cite{kou17} systems.

A simple interpretation reveals the connection between the dynamic heterogeneity and the non-Gaussian diffusion by modeling the heterogeneity with the random instantaneous diffusivity $D^{(t)} $\cite{wang09,wang12}. $P(x,t)$ is, hence, a convolution over $D^{(t)}$ by
\begin{equation}
P(x,t)=\int dD^{(t)}\; G(x,t\vert D^{(t)})P(D^{(t)}), 
\end{equation}
where $G(x,t\vert D^{(t)})$ is the Gaussian kernel for the short segment with the given $D^{(t)}$. 
Chubynsky and Slater \cite{slater14} constructed the dynamics by setting the diffusivity itself as an Ornstein-Uhlenbeck process, which is, hence, temporally correlated. The non-Gaussian behavior exists in the correlation time scale. The recent studies on the diffusion with fluctuating diffusivity \cite{cherstvy16,akimoto16pre,aurell17,sposini18,slezak18} have largely improved our understanding on the non-Gaussian diffusion in the annealed disordered environments where the relaxation time of the environments is assumed on the same scale of the particle diffusion and no spatial structure is considered. 

The annealed assumption may fail, however, when the disordered environments are greatly influenced by the large structures in the media, such as the actin under the cell membrane \cite{andrews08,sungkaworn17} or endoplasmic reticulum in the cytoplasm \cite{li15}. In the case that the structures fluctuate quite slowly, the disordered sample is quasi-static \cite{munder16,jeon16,ghosh15,li15,ye12,zhao16} over the whole experiment. In such a case, the fluctuating diffusivity is correlated in space but not in time. To investigate the non-Gaussian diffusion in the static disordered systems, we have recently constructed a spatially correlated random landscape by a trick from extreme statistics \cite{luo16,luo18}. Employing the trap dynamics with effective temperature $\mu=1$, the local diffusivity on the extreme landscape follows the exponential distribution $P(D^{(l)}/D_0=D)=\exp(-D)$. The exponential tail of $P(x,t)$ is hence expected for small $t$. The model study has revealed a localization mechanism in the ensemble of trajectory segments, which is universal for the quenched disordered cases. Similar phenomena have been already observed in experiments. 

In this paper, we further investigate the trap dynamics on the ``extreme landscape'' of different heterogeneous levels, which are modulated by the effective temperature $\mu$. A coarse-graining (CG) process is introduced to handle the spatial correlation in the landscape via which the equivalence between the current model and the quenched trap model (QTM) \cite{haus87,bouchaud90,barkai11,luo14} are revealed in the long time limit. The extreme landscape is hence a generalization of the QTM, in which the finer structures appear as the spatially correlated
local diffusivity. The non-Gaussian distribution of displacement is, hence, expected on the correlated length scale. As the same in the QTM, subdiffusion due to strong heterogeneity also arises in the current model for $\mu<1$. In the quenched case, a localization happens in the ensemble of the trajectory segments. A peak in $P(x,t)$ around $x=0$ arises accordingly, which is significantly split from the stretched exponential tail. 

The paper is organized as follows. In Sec. \ref{sec_model}, we introduce the extreme landscape and show how the effective temperature of the trap dynamics controls the heterogeneity. In Sec. \ref{sec_cg}, we introduce a coarse-graining process, which connects the current model to the traditional QTM. In Sec. \ref{sec_gaussian}, we investigate the structure of the non-Gaussian distribution of displacement. Section\ref{sec_ds} discusses the biological implication of the results and their connections with other works. A short summary follows in Sec. \ref{sec_sum}.

\section{The trap dynamics on the extreme landscape}
\label{sec_model}

We consider the random walk on a static disordered landscape $\{V_i\}$ in a two-dimensional cubic lattice, 
where $i$ denotes the lattice site. The landscape was proposed to record the information of local minima of random auxiliary landscapes. It can be called the extreme landscape. 
The generation of the extreme landscape $\{V_i\}$ typically follows two steps:
\begin{enumerate}
\item Generate an auxiliary uncorrelated random landscape $\{U_i\}$, 
following the exponential distribution 
\begin{equation}
P(U_i=U)=U_0^{-1}\exp\left(U/U_0\right),\text{\hspace{2em}} U<0.
\end{equation}

\item Assign $V_i$ by the local minimum of $\{U_i\}$ in the $r_c$-neighbourhood of site $i$, i.e., 
\begin{equation}
\label{eq_minrule}
V_i=\min \left\{U_j\vert r_{ij}<r_c\right\}.
\end{equation}

\end{enumerate}
Noting that the auxiliary landscape $\{U_i\}$ is uncorrelated,  $P(V_i=V)$ converges to the limit distribution of extreme statistics for $r_c^2\gg1$, which distribution is known as the Gumbel distribution 
\begin{equation}
\label{eq_gumbel}
P(V_i=V)=\exp\left[V-V_0-\exp\left(V-V_0\right)\right]. 
\end{equation}
The extreme landscape is essentially determined by spatial distribution of the local minima of the auxiliary landscape. Each minimum dominates a range of the neighbour traps, which shape a basin of radius $r_c$ in the extreme landscape. The extreme landscape is constituted by the overlapped extreme basins (See Fig. 1 in Ref. \cite{luo18}). It is, hence, locally correlated up to $2r_c$. 

\begin{figure}
\centering
\includegraphics[width=8.6cm]{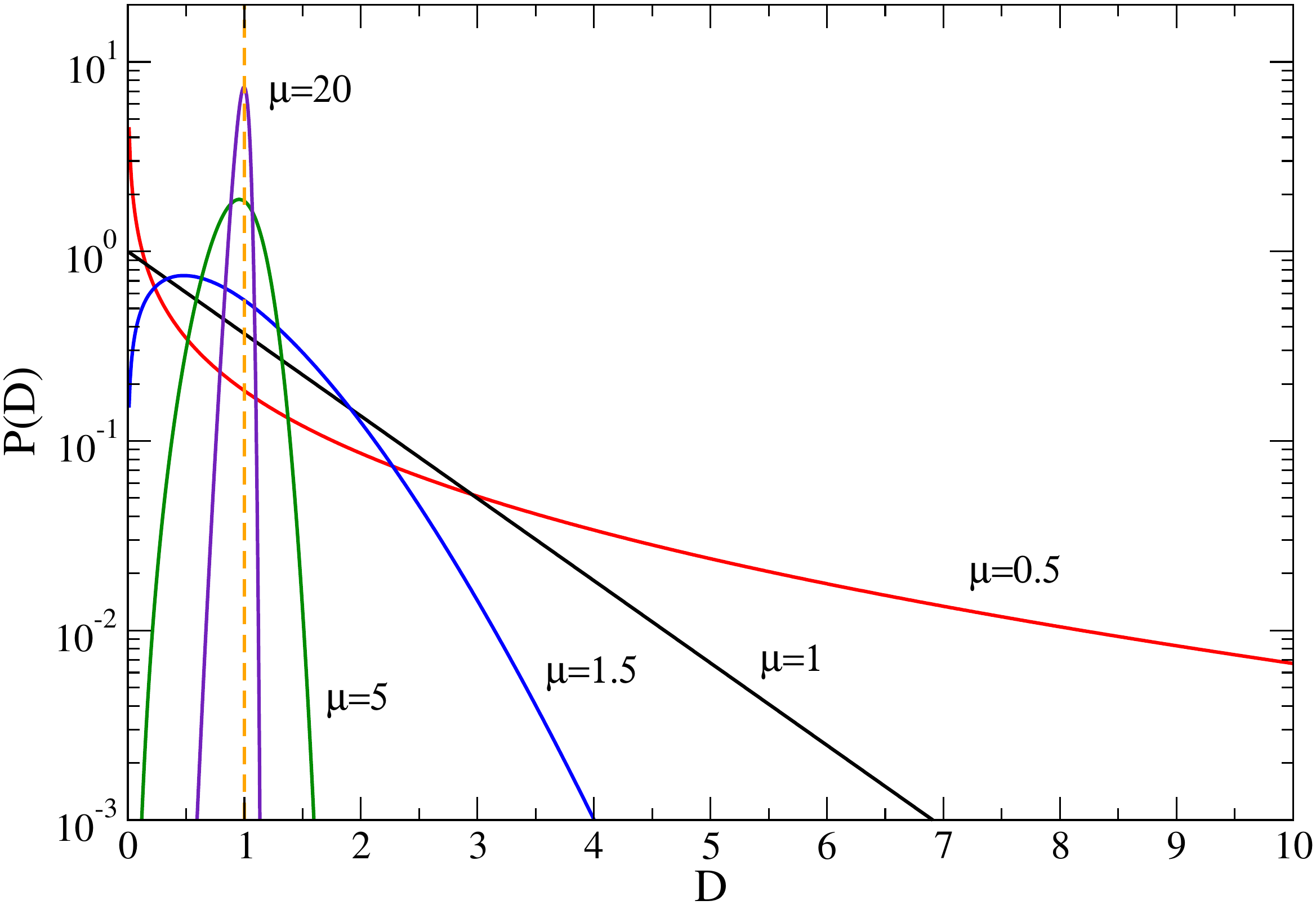}
\caption{\label{fig_pd} The distribution of local diffusivity for various $\mu$'s, given by Eq. (\ref{eq_gamma}). }
\end{figure}

In this paper, the trap dynamics is employed for the random walk on the extreme landscape. 
The escaping rate from the trap $i$ is determined by the trap depth $V_i$ by
\begin{equation}
w_i=w_0\exp(V_i/\mu), 
\end{equation}
where $V_i<0$ for traps and the dynamical parameter $w_0$ gives the timescale. The effective temperature $\mu$ controls the roughness of the landscape and, hence, the spatial heterogeneity of the dynamics. 
The typical sojourn time in trap $i$ can be estimated by 
\begin{equation}
\label{eq_taul}
\tau_i=w_i^{-1}=w_0^{-1}\exp(-V_i/\mu).
\end{equation}
In trap dynamics, the particle at site $i$ jumps to all the nearest-neighbour sites $j$ with even rate $w_{i\rightarrow j}=n_c^{-1}w_i$, where $n_c=4$ is the coordination number in the square lattice. 
The local diffusivity at site $i$ can be hence defined as
\begin{equation}
\label{eq_dl}
D^{(l)}_i\equiv\frac{a^2}{4\tau_i}=\frac{w_0a^2}{4}\exp(V_i/\mu),
\end{equation}
Noting that $\{V_i\}$ follows the Gumbel distribution given by Eq. (\ref{eq_gumbel}), one can see $\{D^{(l)}_i\}$ follows the generalized Gamma distribution with a stretched exponential tail
\begin{equation}
\label{eq_gamma}
P(D')=\mu D'^{\mu-1}\exp(-D'^\mu),
\end{equation}
where $D'\equiv D^{(l)}/D_0$ is scaled by $D_0=w_0a^2\exp(V_0/\mu)/4$. 
Noting $V_0<0$, one can see $D_0$ vanishes in the low temperature cases with $\mu\ll1$, where the walk in the media is frozen. 
To exclude the freezing effects and to focus on the spatial heterogeneity, we rescale the landscape in this paper by setting $D_0=1$. The mean value of $D^{(l)}$ then moderately depends on $\mu$ by $\left<D^{(l)}\right>=\mu^{-1}\Gamma(\mu^{-1})$, where $\Gamma(\cdot)$ is the Gamma function. For intuition, $0.88<\left<D^{(l)}\right><2$ for any $\mu>0.5$. 
Figure \ref{fig_pd} shows the distribution of the rescaled local diffusivity, $P(D^{(l)})$, for some typical temperatures. 
In the high temperature limit, $\mu\rightarrow\infty$, $P(D^{(l)}_i/D_0=D)$ converges to a peak around $D=1$. The dynamics, hence, degenerates to the normal Brownian motion in the homogeneous media. For $\mu=1$, Eq. (\ref{eq_gamma}) turns to $P(D_i^{(l)}/D_0=D)=\exp(-D)$, which has been previously studied \cite{luo18} as a special case of non-Gaussian diffusion with the exponential tail. 

\section{The coarse-graining process and the long-time behavior}

\label{sec_cg}

\begin{figure}
\centering
\includegraphics[width=8.6cm]{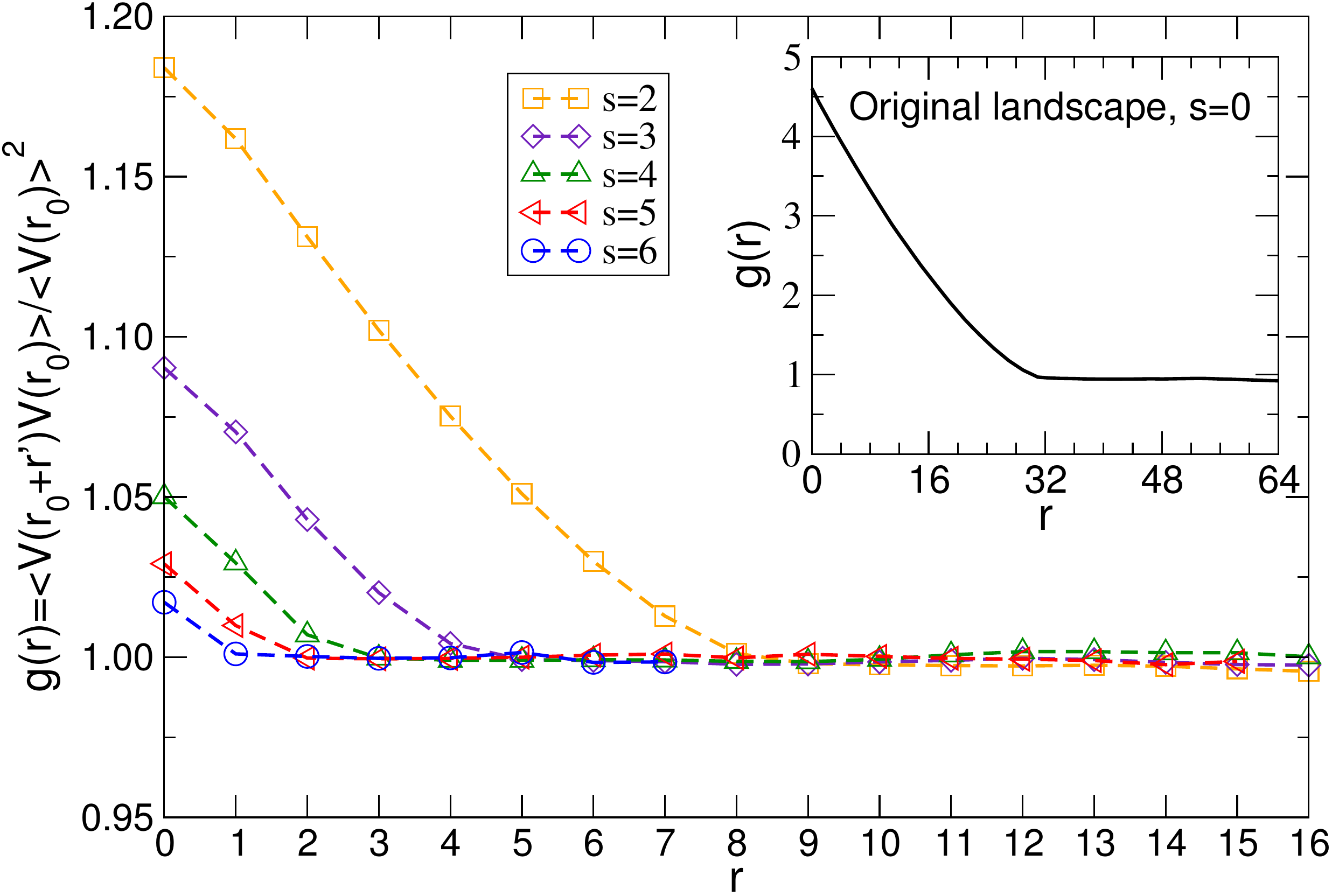}
\caption{\label{fig_vcorr} The pair correlation function $g(r)$ of the coarse-grained landscapes at various CG levels $s$. 
The radius of the extreme basin is set by $r_c=16$. The inset shows $g(r)$ of the original extreme landscape. }
\end{figure}

In this section, we introduce the CG process for the trap model to handle the local correlation in the landscape. 
Considering a sample of finite size and periodic boundaries, the random walk can scan all the traps of the sample 
in the long time limit. The diffusion process achieves a steady state, of which the mean squared displacement (MSD)
$\left<\vert \Delta x(t)\vert^2\right>=\left<\vert x(t)-x(0)\vert^2\right>$ can be written by
\begin{equation}
\left<\vert \Delta x(t)\vert^2\right>=4 D_{\text{dis}}t, \text{ for } t\rightarrow\infty. 
\end{equation}
One can show in trap dynamics that the diffusion coefficient $D_{\text{dis}}$ depends on the mean sojourn time\cite{haus87, bouchaud90,akimoto16} by
\begin{equation}
\label{eq_dt}
D_{\text{dis}}=a^2/4\overline{\tau},
\end{equation}
where the mean sojourn time $\overline{\tau}$ averages over the traps in the sample by
\begin{equation}
\label{eq_taubar}
\overline{\tau}=\frac{1}{N}\sum_{i=1}^N \tau_i. 
\end{equation}
Sharing the spirit with Machta's early work \cite{machta83} on the QTM, we regroup the summands in Eq. (\ref{eq_taubar}) by blocks of neighbours. It leads the CG operation as follows:
\begin{enumerate}
\item In a lattice of $N$ sites, we replace each $2\times2$ block by a single site on a lattice of $N'=N/4$ sites and the lattice constant $a'=2a$. 

\item To keep the sum of all the $\tau_i$'s invariant, the typical sojourn time $\tau'_{q}$ in a coarse-grained site $q$ is set to the sum of those in the original block, 
\begin{equation}
\tau'_q=\sum_{j\in{\text{ block }}q}\tau_{j},
\end{equation}
where $\tau_j$ is the typical waiting time of the $j$th site in block $q$. 
\end{enumerate}
Repeating the operation for $s$ times, we achieve a landscape of CG level $s$, which is constituted by $N^{(s)}=N/4^s$ traps. The mean sojourn time of the coarse-grained landscape is given by
\begin{equation}
\label{eq_tauqbar}
\overline{\tau^{(s)}}=\frac{1}{N^{(s)}}\sum_{q=1}^{N^{(s)}}\tau^{(s)}_q=4^s \overline{\tau},
\end{equation}
where $\tau^{(s)}_q$ denotes the typical sojourn time in the $q$th trap. 
Noting that the lattice constant $a^{(s)}=2^s a$, we see the diffusion coefficient ${D^{(s)}_{\text{dis}}}={(a^{(s)})}^2/(4\overline{\tau^{(s)}})$ is invariant over coarse-graining. 
On the other hand, the spatial correlation quits the coarse-grained landscape, as shown by the pair correlation function of the effective landscape $\{\tilde{V}^{(s)}_q\equiv-\ln\tau^{(s)}_q\}$ in Fig.\ref{fig_vcorr}. One can clearly read the decline of the correlation, which vanishes when the grain size $l=2^s$ is larger than the diameter of the extreme basin $2r_c$. 

\begin{figure}
\centering
\includegraphics[width=8.6cm]{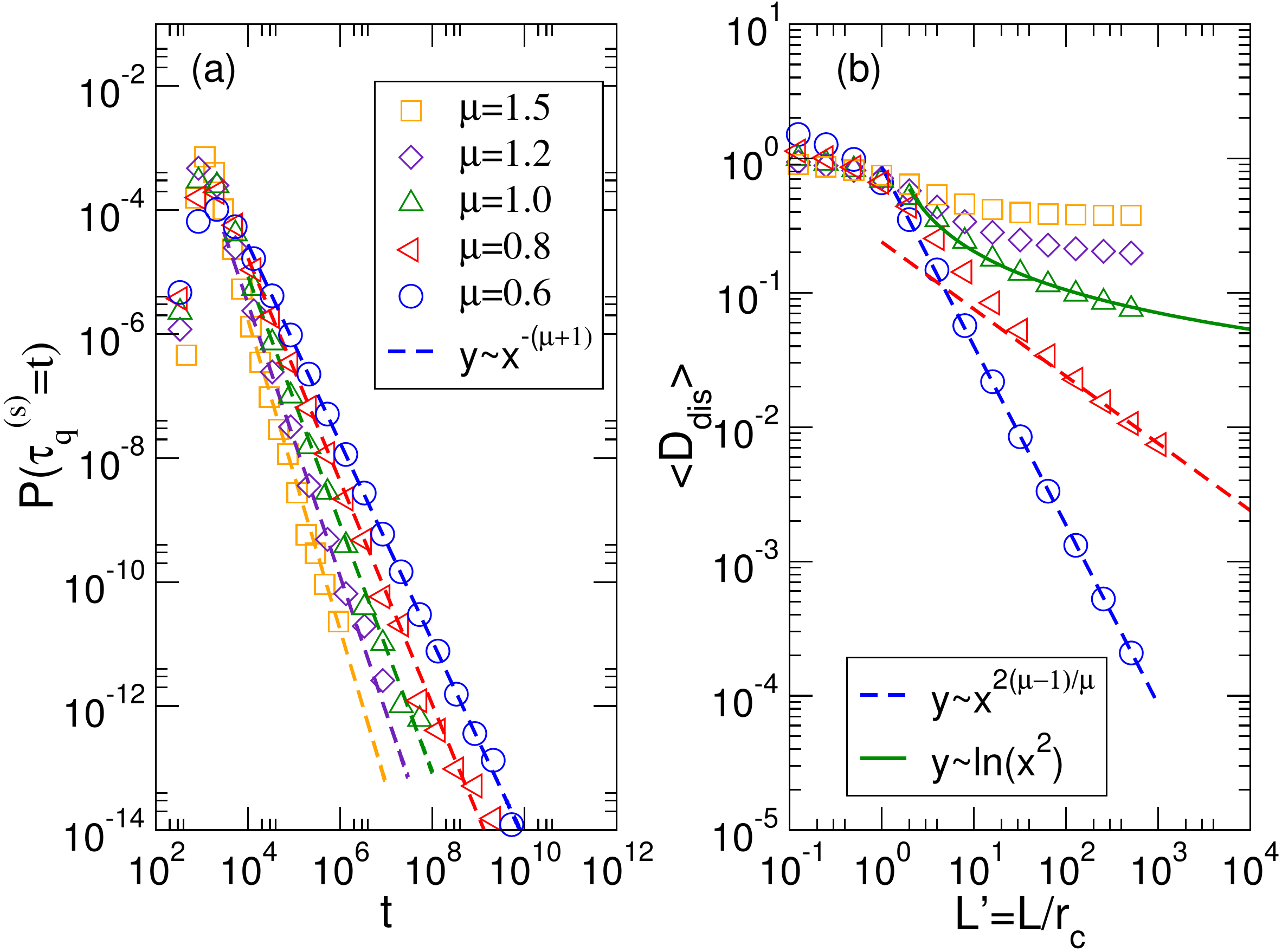}
\caption{\label{fig_taucg} (a) The probability density function of the sojourn time in the traps of the coarse-grained landscape with the CG level $s=6$ from simulations for various $\mu$'s. (b) The size dependence of $\left<D_{\text{dis}}\right>$, where the lattice size is rescaled by the correlation length $r_c$. The symbols denote the numerical results for various $\mu$'s as the same in (a).
}
\end{figure}

Figure \ref{fig_taucg}(a) shows the probability density function of the typical sojourn time in traps $\{\tau_q^{(s)}\}$ of the fully coarse-grained landscapes. One can clearly read the power-law tails contributed by the sojourn time in the deepest traps. Being more precise, the depth of the original traps follows the Gumbel distribution given by Eq. (\ref{eq_gumbel}), of which the tail is merely exponentially shaped. The exponential tail of $P(V_i)$ leads to the power-law tail of sojourn time distribution.
It recalls to us the intensively studied QTM with no spatial correlation, in which the heavy-tailed sojourn time distribution leads to subdiffusion. 

subdiffusion does arise in trap dynamics on the extreme landscape when $\mu\le1$. 
In the rest of the section, we characterize the subdiffusive behavior by the size dependence of diffusion coefficient $D_{\text{dis}}$ of the extreme landscape in a brief way. For more technical details, one can go to the classical reviews\cite{haus87,bouchaud90} and also the recent papers\cite{luo14,luo15,akimoto16}. 

subdiffusion refers to the sublinear time dependence of MSD, 
where $D_{\text{dis}}$ vanishes as the particle scans a broader range of the sample. 
The QTM captures the feature of subdiffusion by the size dependence of $D_{\text{dis}}$, 
which is connected to the mean sojourn time of the sample via Eq. (\ref{eq_dt}). 
For simplicity, we consider the fully coarse-grained landscape with $M=N^{(s)}=N/4^{s}$ traps, where $\{\tau_q^{(s)}\}$ is independently and identically distributed. In the case that the distribution of $\tau_q^{(s)}$ is with a power-law tail, $P(\tau_q^{(s)}=t)\sim c t^{-(\mu+1)}$, one may note a random energy model- (REM-) like transition \cite{derrida81,bouchaud97} happens for $\mu<1$, where the summation of $\tau_q^{(s)}$ in Eq. (\ref{eq_tauqbar}) is dominated by the largest summand. Including more terms in the summation, the typical value of the largest $\tau_q^{(s)}$ increases as $\tau_{\text{typ}}\sim M^{1/\mu}$, which is faster than linear. 
The mean sojourn time $\overline{\tau^{(s)}}$, hence, diverges for large $M$. The vanishing $D_{\text{dis}}$ becomes the consequence.
The generalized central limit theorem suggests the rescaled summation of $\tau_q^{(s)}$ follows the one-sided L\'{e}vy stable distribution by
\begin{equation}
\frac{A}{M^{1/\mu}}\sum_{q=1}^{M}\tau_q^{(s)}\equiv\tilde{\tau}\sim L_{\mu}, \text{ for }\mu<1,
\end{equation}
where the normalization parameter $A$ depends on $\mu$ and $c$.  
Noting $\overline{\tau^{(s)}}=A^{-1}M^{\frac{1-\mu}{\mu}}\tilde{\tau}$ and also ${D_{\text{dis}}}=(a^{(s)})^2/(4\overline{\tau^{(s)}})$, one can estimate the mean diffusion coefficient averaged over samples
by 
\begin{equation}
\label{eq_d1scaling}
\left<D_{\text{dis}}\right>=M^{1-\frac{1}{\mu}}\frac{({a^{(s)}})^2}{4}A\left<\tilde{\tau}^{-1}\right>, 
\end{equation}
and the sample-to-sample fluctuation by
\begin{equation}
\label{eq_d2scaling}
\left<\vert D_{\text{dis}}-\left<D_{\text{dis}}\right>\vert^2\right>=M^{2-\frac{2}{\mu}}\frac{({a^{(s)}})^4}{16}A^2\left[\left<\tilde{\tau}^{-2}\right>-\left<\tilde{\tau}^{-1}\right>^2\right], 
\end{equation}
where the negative moments\cite{chechkin09} of $\tilde{\tau}$ depend only on $\mu$. 
Noting $M=(L/2^s)^2$, one can see $\left<D_{\text{dis}}\right>\propto L^{2(\mu-1)/\mu}$ for $\mu<1$.
In the marginal $\mu=1$ case, the logarithmic size dependence, $\left<D_{\text{dis}}\right>\propto 1/\ln L^2$, 
has been reported in the previous work\cite{luo18}. 
For higher $\mu$, the mean value of $\tau_q^{(s)}$ exists. Self-averaging can be achieved in large samples. $\left<D_{\text{dis}}\right>$ hence converges to a finite value. In other words, the random walk in less heterogeneous landscapes with $\mu>1$ return to the normal Brownian motion in the long time limit. Our simulation confirms the above results on size dependences of $D_{\text{dis}}$ for various $\mu$'s, as shown in Fig.\ref{fig_taucg}(b).

\section{Non-Gaussian diffusion with the stretched exponential tail and the peak around $x=0$}
\label{sec_gaussian}

In this section, we investigate the distribution of displacement of random walk on the extreme landscape.
In practice of data analysis, the distribution of displacement $P(x,t)$ is usually obtained by counting the head-to-tail displacement $x$ of trajectory segments of time duration $t$. 
In this paper, we generate the trajectories by the kinetic Monte Carlo simulation\cite{gillespie77}. To simulate the fully equilibrium case, the initial positions of the walk are randomly generated following the Boltzmann distribution, i.e. $P_i\propto\tau_i=a^2/4D^{(l)}_i$. 
One can find more simulation details in Appendix \ref{app_b}. 



\begin{figure}
\centering
\includegraphics[width=8.6cm]{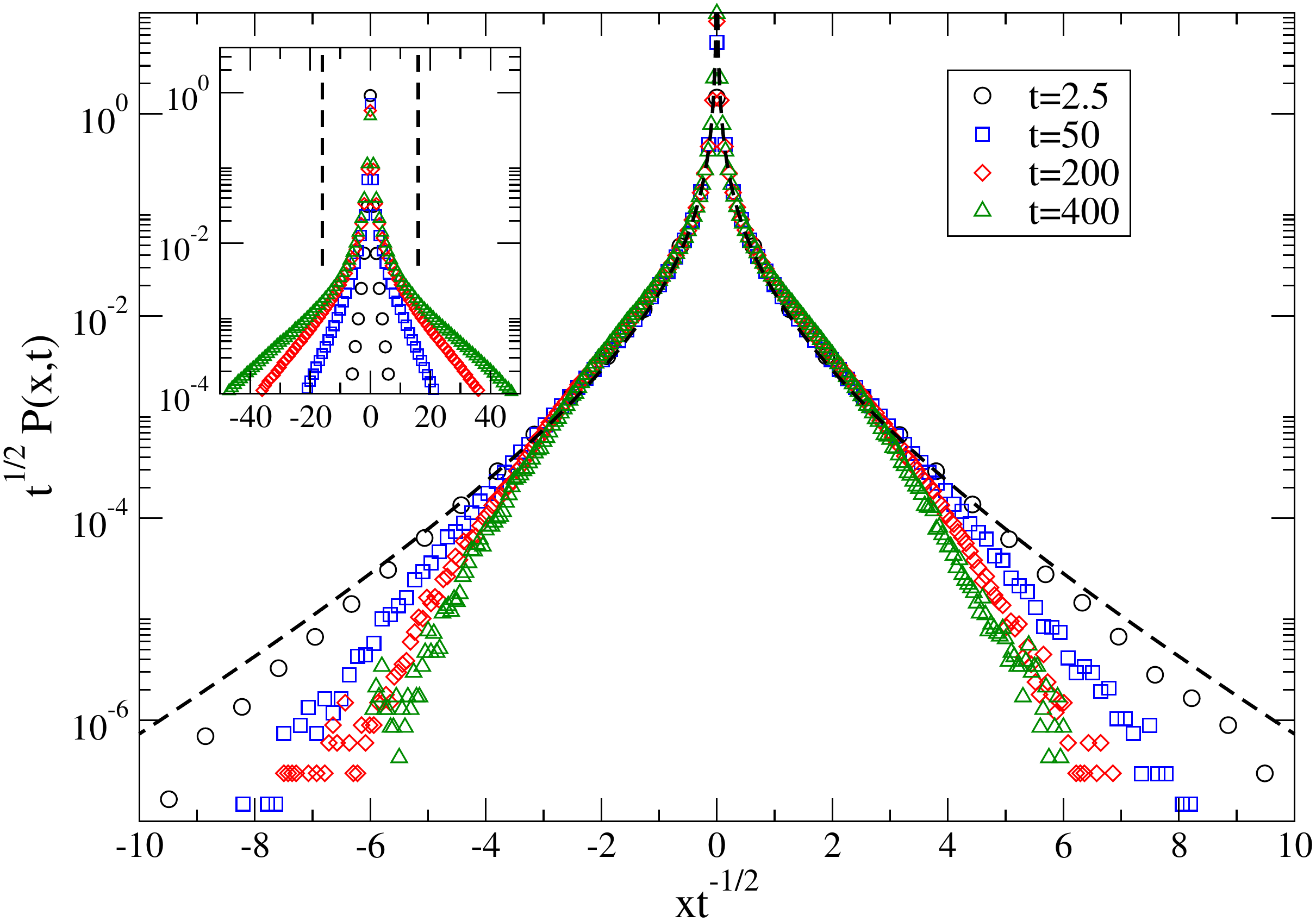}
\caption{\label{fig_px08} $P(x,t)$ for $\mu=0.8$ in the scaled fashion. The symbols are obtained from simulations and the black dashed lines are given according to Eq. (\ref{eq_pdx}). The inset shows the non-scaled distribution, where the radius of the extreme basin $r_c=16$ is marked by the dashed lines for guidance.}
\end{figure}

\begin{figure}
\centering
\includegraphics[width=8.6cm]{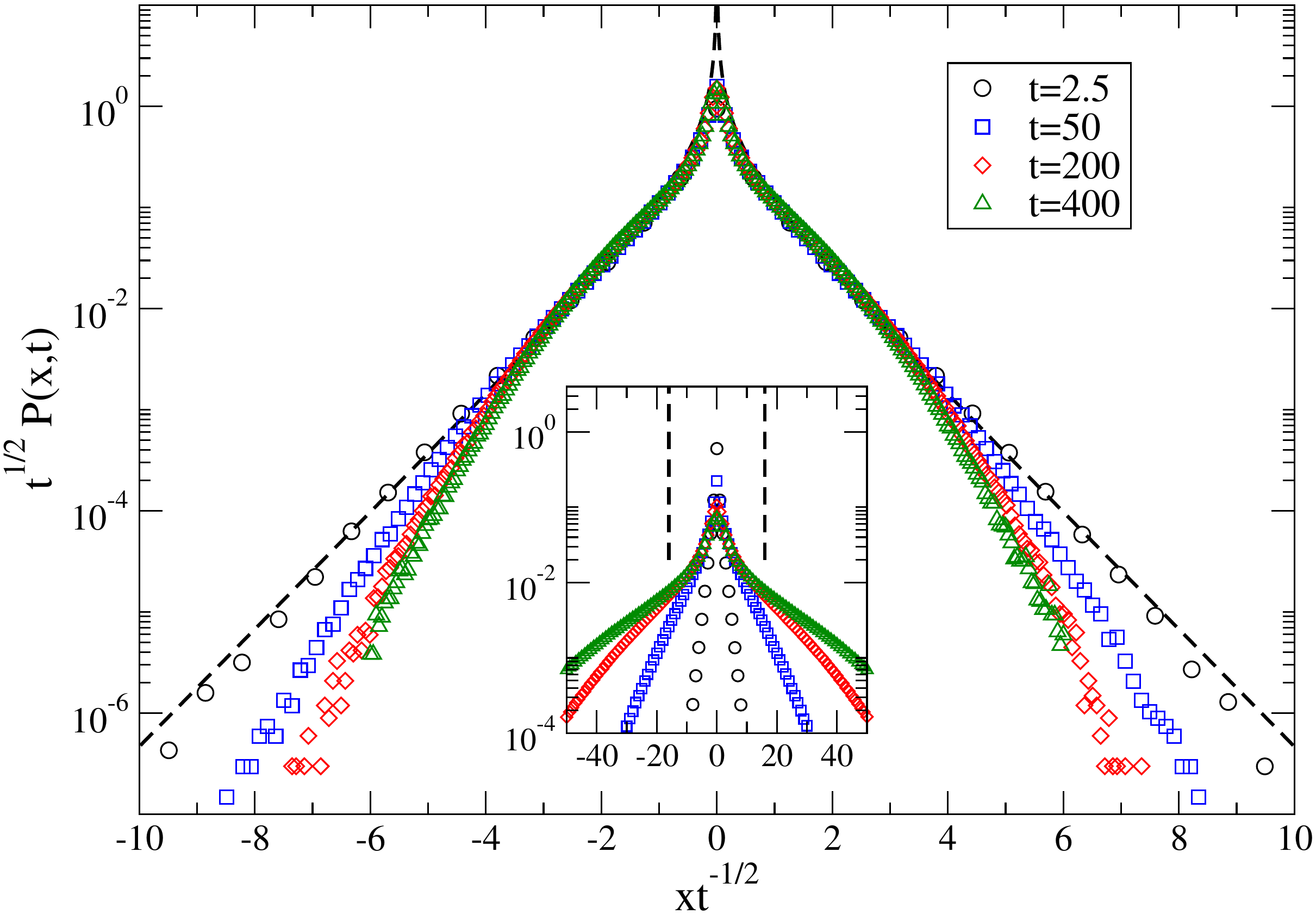}
\caption{\label{fig_px12} Same as Fig.\ref{fig_px08}, but for $\mu=1.2$.}
\end{figure}

\begin{figure}
\centering
\includegraphics[width=8.6cm]{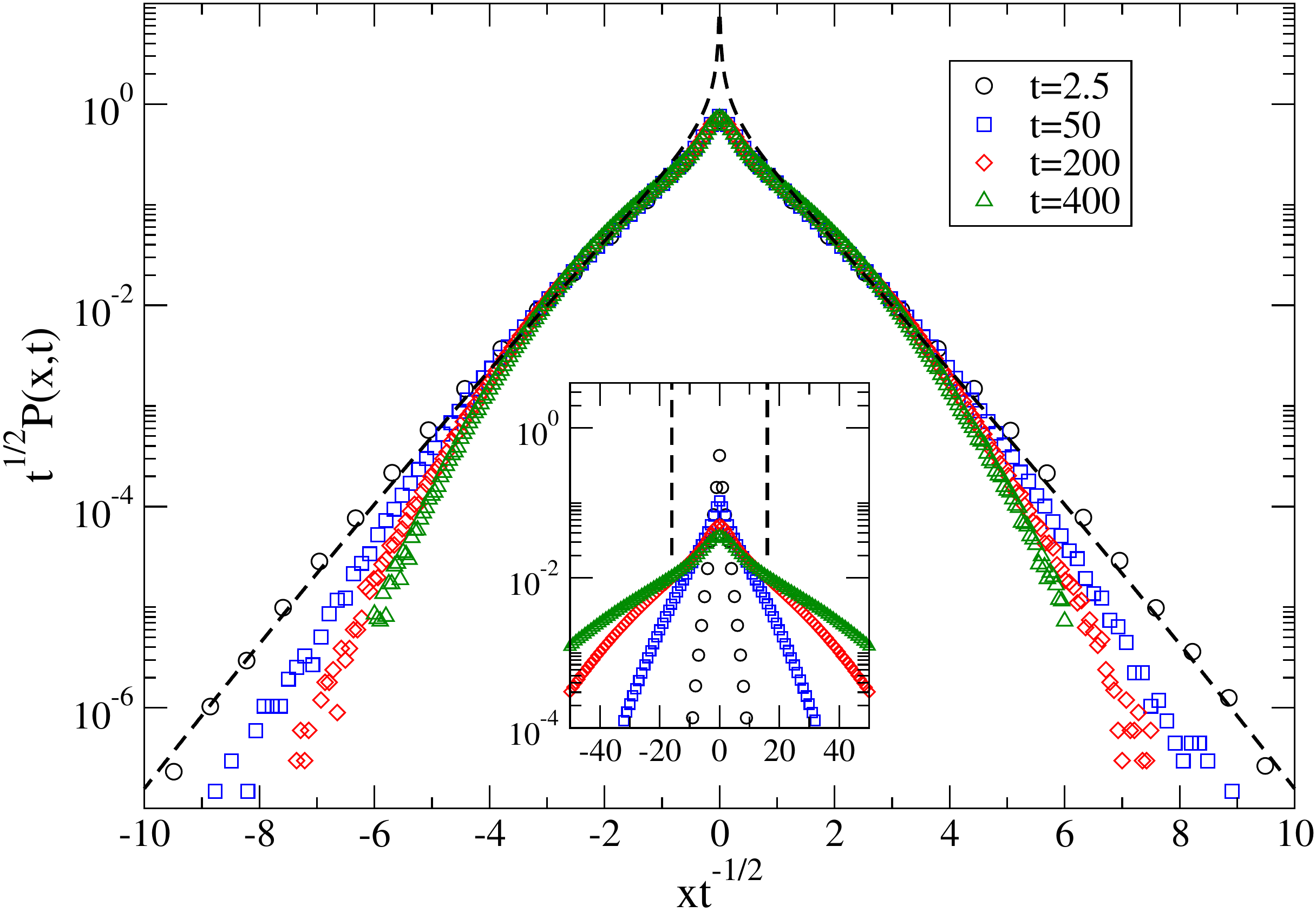}
\caption{\label{fig_px15} Same as Fig.\ref{fig_px08}, but for $\mu=1.5$.}
\end{figure}

The main results are shown in Fig. \ref{fig_px08}-\ref{fig_px15}, where $P(x,t)$'s are presented in the scaled style for various $\mu$'s and $t$'s. The distribution for small $t$ can be estimated by the superstatistic assumption where the instantaneous diffusivity $D^{(t)}$ of the short segments can be approximated by the local diffusivity $D^{(l)}$ of the extreme basin as long as the time is too short for the particle to leave the original basin. Given the initial site $i$ of a short segment, one can expect the displacement of the segment follows the Gaussian distribution governed by a single diffusivity $D^{(t)}=D^{(l)}_i$ as
\begin{equation}
G(x,t\vert D^{(t)})=\frac{1}{\sqrt{4\pi D^{(t)}t}}\exp\left(-\frac{x^2}{4D^{(t)}t}\right).
\end{equation}
Counting all the segments of various $D^{(t)}$'s, $P(x,t)$ follows a convolution
\begin{equation}
\label{eq_conv}
P(x,t)=\int_0^{\infty}dD^{(t)}\; G(x,t\vert D^{(t)})P(D^{(t)}\vert D_{\text{dis}}), 
\end{equation}
where $P(D^{(t)}\vert D_{\text{dis}})$ is the distribution of the instantaneous diffusivity, 
recording the local diffusivity of the extreme basin visited by each short segment. 
Noting in the equilibrium state the segments sample the landscape with the Boltzmann weight, one can see 
\begin{equation}
\label{eq_marginal}
P(D^{(t)}=D\vert D_{\text{dis}})=\sum_i P(D^{(l)}_i=D\vert D_{\text{dis}})P(x_i\vert D_i^{(l)},D_{\text{dis}}),
\end{equation}
where $P(x_i\vert D_i^{(l)},D_{\text{dis}})$ is the Boltzmann weight of trap $i$ in the sample with $N$ traps. 
Employing Eq. (\ref{eq_dl}), (\ref{eq_dt}), and (\ref{eq_taubar}), it can be explicitly written by
\begin{equation}
\label{eq_boltzmann}
P(x_i\vert D_i^{(l)},D_{\text{dis}})=\frac{\tau_i}{\sum_{j=1}^{N}\tau_j}=\frac{D_{\text{dis}}}{ND^{(l)}_i}. 
\end{equation}
Noting Eq. (\ref{eq_gamma}), one can see Eq. (\ref{eq_conv}) is indeed a convolution of the generalized Gamma distribution and the Gaussian distribution. Sposini {\it et al.}\cite{sposini18} offer several approaches for the estimation of the convolution. 
In Appendix \ref{app_a}, a saddle point approach is introduced, which gives the correct large-$x$ asymptotic behavior by
\begin{equation} 
\label{eq_pdx}
P(\tilde{x},t)\approx \frac{1}{\sqrt{4t}}AD_{\text{dis}}\tilde{x}^{(\mu-3)/(\mu+1)}\exp\left[-B\tilde{x}^{2\mu/(1+\mu)}\right],
\end{equation}
where $\tilde{x}=\sqrt{x^2/4t}$ and the prefactors $A$ and $B$ depend only on $\mu$. 
For the $\mu=1$ case, it returns to the simple expression
$P(x,t\vert D_{\text{dis}})=D_{\text{dis}}x^{-1}\exp\left(-x/\sqrt{t}\right)$, which has been obtained in Ref. \cite{luo18}. 
For the $\mu\neq1$ cases, the stretched or shrunk exponential tail is suggested by Eq. (\ref{eq_pdx}). It is well confirmed by the simulation with $t=2.5$, as shown by the dashed lines in the figures. The deviance arises for $\mu<1$ at the end of the tail (see Fig.\ref{fig_px08}) in which in the frozen case most segments are localized in the deepest traps and too few segments scan the most mobile region. The statistics error becomes significant for the rare ``mobile'' events, which are only $1/10^{7}$ of all the segments. 

It is a bit surprising that the asymptotic expression [Eq. (\ref{eq_pdx})] works even to small $x$, where a peak $P(x)\sim(x^2/4t)^{(\mu-3)/(2\mu+2)}$ is expected. The peak is mainly contributed by the segments in the deepest traps of the sample, which are heavily weighted in the ensemble of segments. Noting the constraint $P(x_i\vert D_i^{(l)},D_{\text{dis}})<1$, one can learn from Eq. (\ref{eq_boltzmann}) that the local diffusivity in a given sample is bounded by $D^{(l)}>D_c\equiv D_{\text{dis}}/N$. The height of the peak at $x=0$ is hence also bounded,  which can be estimated as
\begin{equation}
\label{eq_ph}
P(x=0,t\vert D_{\text{dis}})\approx\frac{D_{\text{dis}}}{\sqrt{4\pi t}}\Gamma\left(\frac{2\mu-3}{2\mu},D_c^{\mu}\right).
\end{equation}
Here $\Gamma(\alpha,z)=\int_z^{\infty}dt\;t^{\alpha-1}\exp(-t)$ is the incomplete Gamma function. 
In the case with $\mu<3/2$, $\Gamma\left((2\mu-3)/2\mu,D_c^{\mu}\right)\sim D_c^{\mu-3/2}$ for small $D_c$, which diverges when $D_c\rightarrow0$. It is interesting to note that $P(D^{(l)}=0)=0$ for $\mu>1$ (see Fig.\ref{fig_pd}). The peak appearing in the $1<\mu<3/2$ cases is purely from the localization in the ensemble of trajectory segments, which is a unique phenomenon in the static disordered media. 

The superstatistics assumption is not suitable for longer $t$ in which case the particle may visit multiple extreme basins. Averaging over various $D^{(l)}$'s of the basins, the tail of $P(x,t)$ for longer $t$ gradually deviates from Eq. (\ref{eq_pdx}) as shown in the figures. On the other hand, the peak contributed by the particles localized in the deepest basins relaxes in different timescales. 
The sharp peak persists for a very long time in the subdiffusive $\mu=0.8$ case, since a genuine glass transition (REM-like transition) drives the deepest trap away from the others. The waiting time in the deepest trap is magnitudes larger than that in the others. In the diffusive $1<\mu<3/2$ cases, the peak due to the localization in the ensemble of segments can still be identified from the tail even for very long $t$, which is more clearly shown by the non-scaled distribution given in the insets of the figures. The whole particle (segment) populations are, hence, split into the ``mobile'' and ``immobile'' states until the  localized particle eventually escapes from the extreme basin of the deepest traps. The size of the extreme basin, hence, gives a length scale separating the peak and the tail, which is marked in the insets of the figures.

\section{Discussion}
\label{sec_ds}

\subsection{Population splitting}
The heterogeneity of the disordered media often introduces different dynamical states in the diffusion process. 
In the model of the aged continuous time random walk \cite{barkai2003prl,barkai2003jcp,schulz13,schulz14,song18}, a portion of particles is localized, which contributes a peak around $x=0$ in $P(x,t)$ due to the heavy-tailed waiting time. The phenomenon of ``population splitting'' is, hence, reported where the displacements of the ``immobile'' and ``mobile'' particles are well split. The recent simulation \cite{jeon16} and experiment\cite{cherstvy19} reports that the position-dependent heterogeneity can also split the particles into subgroups of different dynamical features. It could be understood by noting the particle (or segment) reports only the local dynamical properties as long as the single trajectory has not scanned the whole sample. The long-displacement tails are hence piecewise-fitted according to the subgroups in such cases. 

In this paper, $P(x,t)$ is naturally constituted by the non-Gaussian tail and the peak around $x=0$ out of a different origin - the localization in the ensemble of segments. The tail and the peak split for large $t$, which can be roughly identified as the ``immobile'' and ``mobile'' states. In the typical experimental operation time, most particles can only scan the local environment. Strong fluctuations even survive from averaging the segments along each trajectory, whereas self-averaging is absent in trajectories. The self-averaging can be quantified by the trajectory-to-trajectory fluctuation of time-averaged mean square displacement (TAMSD) as a function of observing time and the ergodicity breaking (EB) parameter \cite{he08,cherstvy13,metzler14,jeon14,jeon16}. The EB parameter estimated from the simulation data report significant nonequivalence between the statistics of a single trajectory and that of the particle ensemble for both the subdiffusive $\mu<1$ cases and the diffusive $\mu>1$ cases even for the maximum simulation time. One can find more details in Appendix C. 



\subsection{The localization observed\\ in (quasi-)static disordered systems}
We note the localization mechanism is universal for all the static disordered systems, but not limited to the  extreme landscape. Some phenomena have been already observed in the previous experiment and simulation works on cell membrane, cytoplasm, and also in the phonon transport in the disordered oscillator chain. 

{\it Cell membrane.} He~{\it et~al.} tracked the protein embedded in the membrane of a living cell, which is crowded by transmembrane protein and other structures\cite{engelman05,sungkaworn17}. The exponential tails of $P(x,t)$ and $P(D^{(t)})$ were reported\cite{he16}, whereas the sharp peaks around $P(x=0,t)$ and $P(D^{(t)}=0)$ have also been noted. The peak could be a signal of the localization in the membrane environment, which strongly interact with the underlying large structures, such as the actin network and other cytoskeletal cortices, and could, hence, be stable for a long time \cite{sungkaworn17}. In the simulation work by Jeon {\it et al.} \cite{jeon16}, it was shown that, in lack of the underlying large structures, the protein crowding can also freeze the diffusion map in the time scale of lipid motion. The non-Gaussian diffusion of lipids arises in the simulation due to the heterogeneity introduced by the protein matrix which relaxes much slower than the lipids diffuse. The non-Gaussian $P(x,t)$ significantly splits into two pieces, which could be a consequence of the localization in the quasistatic diffusion map.

{\it Cytoplasm.} The non-Gaussian diffusion was reported by Munder {\it et~al.} in cytoplasm\cite{munder16}. The perfect exponential distribution was observed in the cell under a normal condition. In the energy-depleted cells where the cellular ATP is largely reduced, a stretched exponential tail and a sharp peak around $x=0$ appears in $P(x,t)$. It is known that the cell constantly employs active motors to fluctuate the cytoskeleton \cite{guo14}. The energy-depletion removes all the active cellular dynamics, in which case the cytoplasm is less fluid and more glass-like. The sharp peak of $P(x,t)$ could be a clue of the frozen and static disordered cytoplasm in the energy-depleted condition. 

{\it Phonon transport.} The non-Gaussian energy diffusion has been reported by Wang {\it et al.} in their study on the one-dimensional disordered oscillator chain \cite{zhao16}. The sharp peak again appears in the static disordered case as a consequence of Anderson localization. In the aspect of phonon transport, it can be also understood that the phonon is localized in the static disordered chain. 

\subsection{The stretched exponential tail of $P(x,t)$}
The tail of a non-Gaussian distribution of displacement is not necessary being in the exact exponential form. The stretched exponential tail is a more common case. A class of non-Gaussian diffusion with a stretched exponential tail has been investigated by Sposini {\it et al.} \cite{sposini18} in two annealed models where the instantaneous diffusivity follows the generalized Gamma distribution. We show in this paper the generalized Gamma distributions [Eq.\ref{eq_gamma}] can be the direct consequence of the trap dynamics on the extreme landscape. The stretched exponential tail of $P(x,t)$, hence, appears in this case of static disorder. 

One would be more careful to estimate the shape of the non-Gaussian tail in the case of limited statistics, which is a common case in experiments. 
In the presence of the localization in static disorder, the tail and the peak would strongly interfere around the center of $P(x,t)$. The exponent can only be correctly estimated in the genuine tail region for large $x$ where the statistics is usually poor. 
One can also read in Figs.\ref{fig_px08}-\ref{fig_px15} the scaled distribution collapse very well for all the $t$s until the genuine tail region with $t^{1/2}P(x,t)\sim10^{-5}$. We suggest, in the static disordered case, to observe the distribution in the nonscaled way as the insets of the figures. The relaxation of the non-Gaussian tail is expected for $x>r_c$. In more physical words, the self-averaging happens only when the particle leaves the original local structure. 

\subsection{The effective temperature $\mu$\\ and the connection to QTM}
In this theoretical paper, the heterogeneity of the static diffusion map $\{D^{(l)}_i\}$ is modulated by the effective temperature $\mu$, which can be hardly measured in experiments. It is more practical to reconstruct the diffusion map from the trajectories and then to characterize the dynamic heterogeneity by $P(D^{(l)}_i)$, as shown in Fig. \ref{fig_pd}. 
In experiments, there is not a general way to modify the heterogeneity level via a certain parameter, such as $\mu$, whereas keeping the local structure. We propose here two specific experiments. In the colloidal system, one can prepare the static disordered potential by the optical tweezers to control the motion of the colloidal particles. The heterogeneity of the potential can be hence adjusted by the setup of the tweezers. In the experiments on the living cells, one can modulate the crowding level of the cytoplasm via osmotic pressure. The freezing transition from normal diffusion to subdiffusion could be expected under proper experiment setups. 

Finally, we clarify the connection between the extreme landscape and the celebrated QTM. Focusing on the long time limit, a coarse-graining process is introduced to eliminate the local correlation in the landscape. Since the tail of Gumbel distribution can be well approximated by an exponential one, the CG process eventually leads us to the QTM with heavy-tailed waiting time. The QTM with uncorrelated traps has been intensively studied since the early 1980s\cite{machta81,machta83,machta85,haus87}. This successful model helps us understand subdiffusion in static disordered media\cite{bouchaud90,monthus03,bertin03,barkai11,luo14}. The trap dynamics on the extreme landscape is indeed an extension of the QTM of which the local structures introduce the non-Gaussian diffusion. 

\section{Summary}
\label{sec_sum}

To summarize, we have investigated the trap dynamics on the ``extreme'' landscape, of which the heterogeneity can be continuously modulated by the effective temperature $\mu$. We show in the long-time limit, the model is equivalent to the celebrated quenched trap model with no spatial correlation. subdiffusion in the extreme landscape is, hence, expected and confirmed in the low temperature region with $\mu<1$. Our analytical study reveals the connection between the stretched exponential tail of $P(x,t)$ and the dynamic heterogeneity. We note a localization mechanism in the ensemble of segments, due to which the ``immobile'' particles are well split from the ``mobile'' ones. It introduces population splitting even in the diffusive $1<\mu<3/2$ cases, while the subdiffusion is absent. 

The molecules in the ``immobile'' state are believed to play a key role in the biological processes requiring a long reaction time. A recent biology study suggests that the actin structures under the cell membrane help the formation of signaling hot spots on the membrane, where the signaling protein tends to stay and work\cite{sungkaworn17}. In this paper, we show in the disordered media fixed by the large structures, the ``immobile'' state spontaneously appears due to the localization in the ensemble of segments, which is merely a consequence of the equilibrium Boltzmann distribution in the static landscape. It provides a hint that how the cell benefits from the (quasi)static structure of the membrane. 

\begin{acknowledgements}
We would like to thank Hui Li for constant inputs and challenges from the experiment side. This work is supported by National Natural Science Foundation of China (Grant No. 11705064, 11675060,  91730301), Fundamental Research Funds for the Central Universities (Grant No. 2662016QD005), and the Huazhong Agricultural University Scientific and Technological Self-innovation Foundation Program (Grant No.2015RC021). 
\end{acknowledgements}

\appendix

\section{Non-Gaussian tail of $P(x,t)$ for small $t$}
\label{app_a}

In this appendix, we estimate the convolution of Eq. (\ref{eq_conv})
\begin{equation}
P(x,t)=\int_0^{\infty}dD^{(t)}\; G(x,t\vert D^{(t)})P(D^{(t)}\vert D_{\text{dis}}), 
\end{equation}
where $P(D^{(t)}\vert D_{\text{dis}})$ is determined by the distribution of local diffusivity
$P(D^{(l)}_i=D\vert D_{\text{dis}})$ and the Boltzmann weight $P(x_i\vert D_i^{(l)},D_{\text{dis}})$ via Eq. (\ref{eq_marginal}). 

We note in any sample, the diffusion coefficient $D_{\text{dis}}$ is determined by the configuration of $\{D^{(l)}_i\}$. 
Known $D_{\text{dis}}$ of the sample, $D^{(l)}$ is hence bounded by $D^{(l)}_i>D_c\equiv D_{\text{dis}}/N$, 
which can be read from the natural constraint $P(x_i\vert D_i^{(l)},D_{\text{dis}})<1$. 
The analysis in the previous work (see Appendix B of \cite{luo18}) suggests the conditional probability $P(D^{(l)}_i=D\vert D_{\text{dis}})$ can be approximated by $P(D^{(l)}_i=D)$ for $D>D_c$, which is provided by Eq. (\ref{eq_gamma}).
Combining also Eqs. (\ref{eq_marginal}) and  (\ref{eq_boltzmann}), we can obtain the distribution of instantaneous diffusivity reported by the short segments,
\begin{equation}
\label{eq_pdt}
P(D^{(t)}_i=D\vert D_{\text{dis}}) \approx
\begin{cases}
0, & D<D_c \\
D_{\text{dis}}\mu D^{\mu-2}\exp(-D^\mu), & D\ge D_c
\end{cases}.
\end{equation}
The explicit expression of Eq. (\ref{eq_conv}) is hence written as
\begin{equation}
\label{eq_conv1}
P(x,t\vert D_{\text{dis}})=\frac{D_{\text{dis}}}{\sqrt{4\pi t}}\int_{D_c}^{\infty}dD\; \mu D^{\mu-5/2}\exp\left(-D^\mu-\frac{x^2}{4Dt}\right).
\end{equation}
One may note the segments with small $D$ hardly contribute to the non-Gaussian tail for $x^2\gg 4D_c t$. In such a case, 
the lower bound of the integral can be released to $D_c=0$. Sposini {\it et al.} \cite{sposini18} have estimated the integral
via the Fox {\it H}-function and other approaches. 
We provide a saddle point approach below, which also gives the correct large-$x$ behavior. 

Let $\tilde{x}=\sqrt{x^2/4t}$, $\tilde{D}=D/\tilde{x}^2$, and $f(\tilde{D})=\tilde{x}^{2\mu}\tilde{D}^\mu+1/\tilde{D}$. The concerned convolution can be written by
\begin{equation}
P(\tilde{x},t\vert D_{\text{dis}})=\frac{D_{\text{dis}}}{\sqrt{4\pi t}}\tilde{x}^{2\mu-3}\int_{0}^{\infty}d\tilde{D}\; \mu \tilde{D}^{\mu-\frac{5}{2}}\exp\left(-f(\tilde{D})\right).
\end{equation}
For $\tilde{x}^2\gg1$, the saddle point approximation suggests
\begin{equation}
P(\tilde{x},t)\approx\frac{D_{\text{dis}}}{\sqrt{4\pi t}}\tilde{x}^{2\mu-3}\mu D_s^{\mu-\frac{5}{2}}\sqrt{\frac{\pi}{f''(D_s)}}\exp\left(-f(D_s)\right),
\end{equation}
where the saddle point $D_s=(\mu)^{-1/(1+\mu)}\tilde{x}^{-2\mu/(1+\mu)}$ gives the minimum value of $f(D)$ by
\begin{equation}
f(D_s)=(1+\mu)\mu^{-\mu/(1+\mu)}\tilde{x}^{2\mu/(1+\mu)}, 
\end{equation}
and
\begin{equation}
f''(D_s)=(1+\mu)\mu^{2/(1+\mu)}\tilde{x}^{6\mu/(1+\mu)}.
\end{equation}
One can hence gets
\begin{equation}
P(\tilde{x},t)\approx \frac{1}{\sqrt{4t}}AD_{\text{dis}}\tilde{x}^{(\mu-3)/(\mu+1)}\exp\left[-B\tilde{x}^{2\mu/(1+\mu)}\right],\end{equation}
where $A=(1+\mu)^{-\frac{1}{2}}\mu^{\frac{2}{1+\mu}}$ and $B=(1+\mu)\mu^{-\frac{\mu}{1+\mu}}$.

\section{The simulation details}
\label{app_b}


The disordered sample is first generated via the two-step
procedure given in Sec. \ref{sec_model}. The size of the samples $\{V_i\}$ is chosen as $L_x=L_y=1024$, whereas the radius of the extreme basin is set as $r_c=16$. 

$10^4$ trajectories are generated on the sample. The initial site of each trajectory is randomly chosen following the Boltzmann distribution by $P_i=\tau_i/\sum_i\tau_i$, to simulate the full equilibrium case. 
Assuming the particle arrives at site $i$ after $k-1$ jumps with the total waiting time $t=\sum_{s=1}^{k-1}t_s$, the following jump on the lattice, $(i,t)\rightarrow(j,t+t_k)$ , is generated by two steps following Gillespie's approach\cite{gillespie77}:
\begin{enumerate}
\item The waiting time $t_k$ in site $i$ is generated following the exponential distribution \begin{equation}
P(t_k)=\tau_i^{-1}\exp\left(-t_k/\tau_i\right). 
\end{equation}
\item The destination of the jump, site $j$, is chosen from the nearest-neighbours of site $i$ by an even rate. 
\end{enumerate}
The simulation is terminated when the total waiting time $t=\sum t_i$ reaches an upper boundary $t_{\text{max}}=5\times10^4$, which can be understood as the finite observation time in experiments. The periodic boundary condition is also applied in the long time simulation. 

$12000$ trajectories are generated for each sample. The simulated trajectories record the waiting time and the direction of each jump, while the trajectories obtained in experiments records the particle position with fixed time interval $\Delta t$. The simulated trajectories are transformed to the experiments style with $\Delta t=2.5$ as the time resolution,  which is discretized into $t_{\text{max}}/\Delta t=20000$ frames. It is a reasonable number for a $10$-min imaging experiment recording $30$ frames per second, whereas the experiment can hardly track $12000$ particles at the same time. 

\FloatBarrier

\section{The fluctuation among trajectories}

\begin{figure}
\centering
\includegraphics[width=8.6cm]{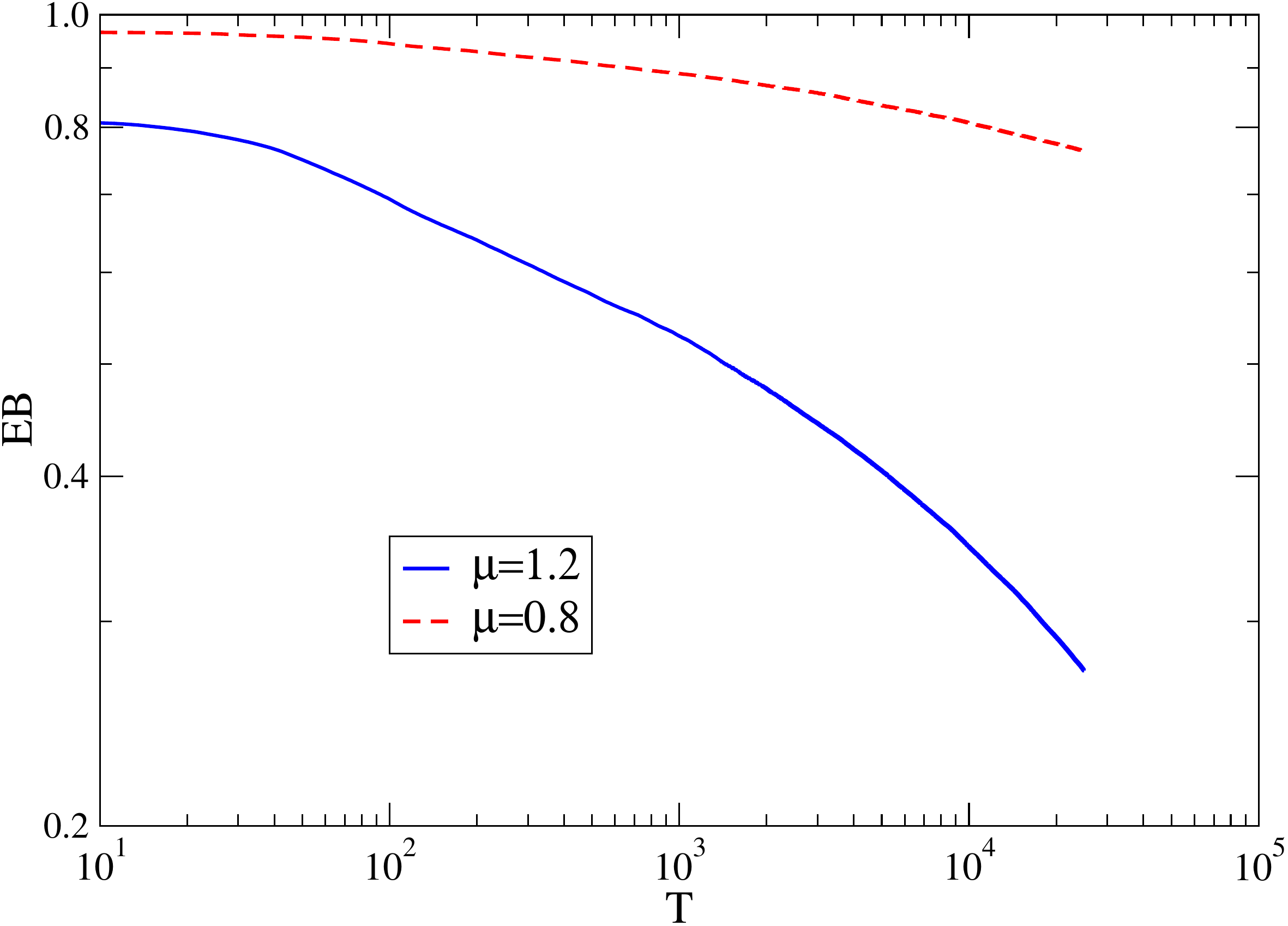}
\caption{\label{fig_eb} The EB parameter as a function of observation time $T$, for $\Delta=50$ in the subdiffusive $\mu=0.8$ case and in the diffusive $\mu=1.2$ case. }
\end{figure}

%

In this appendix, we present the fluctuation among the time-averaged mean square displacement (TAMSD) via the ergodic breaking (EB) parameter. 

TAMSD is defined along a single trajectory as
\begin{equation}
\overline{\delta^2(\Delta,T)}=\frac{1}{T-\Delta}\int_0^{T-\Delta}dt'\;\vert x(t'+\Delta)-x(t')\vert^2,
\end{equation}
where $\Delta$ is the lag time, and $T$ is the observation time along the trajectory. The trajectory-to-trajectory fluctuation can be measured by EB parameter as
\begin{equation}
\text{EB}(\Delta,T)=\frac{\left<\left[\overline{\delta^2(\Delta,T)}\right]^2\right>-\left<\overline{\delta^2(\Delta,T)}\right>^2}{\left<\overline{\delta^2(\Delta,T)}\right>^2}. 
\end{equation}
$\text{EB}=0$ in the full ergodic case, while $\text{EB}=0$ in the fully nonergodic case. Figure \ref{fig_eb} shows EB parameter for $\Delta=50$ and various $\mu$, which decay much slower than the normal Brownian case $\text{EB}\sim\Delta/T$. Significant trajectory-to-trajectory fluctuations persist through the simulation since only few trajectories have span the whole sample in the maximum observation time $t_{\text{max}}$. 


\FloatBarrier


\end{document}